\pdfoutput=1
\documentclass[twocolumn,showpacs,prb]{revtex4}
\usepackage{graphicx}
\usepackage{dcolumn}
\usepackage{bm}

\def\g2{\kappa}
\def\ketbra+-r{ r_p(k_x)\vert{k,p} \rangle^+ \langle{k,p} \vert^- +r_s(k_x) \vert{k,s}\rangle \langle{k,s}\vert}

\begin{document}
\title{ Surface plasmon scattering by shallow and deep surface defects }
\author{Giovanni Brucoli}
\author{L. Mart\'{\i}n-Moreno}
 \affiliation{ Instituto de Ciencia de Materiales de Arag\'{o}n and Departamento de F\'{i}sica de la Materia Condensada,
CSIC-Universidad de Zaragoza, E-50009, Zaragoza, Spain }
\email{gianni@unizar.es}
 \begin{abstract}
Surface plasmon scattering by 1D indentations and protrusions is
examined, mainly in the optical regime. The width of the defects is
fixed, while its height is varied. Both individual and arrays of
defects are considered. Protrusions mainly reflect the incident
plasmons in the optical range. Indentations mainly radiate the
incident plasmon out of plane. An indentation produces maximum
reflections and out-of-plane radiation at the same wavelength, when
its interaction with the incident surface plasmon is resonant.
Protrusions, in general, exhibit maximum reflection and radiation at
different wavelengths.  Shallow arrays of either defects produce a
photonic band-gap, whose spectral width can be broadened by
increasing the defects height or depth. At wavelengths inside the
band-gap  ridge arrays reflect SPPs better than groove arrays, while
groove arrays radiate SPPs better than ridge arrays.
\end{abstract}
 \maketitle
\section{Introduction}
 Surface Plasmon Polaritons (SPPs) are bound modes that confine  light at the interface separating a metal from a
dielectric. The study of surface plasmons is an active research
field sometimes referred to as plasmonics\cite{Plasmonic,Zayats}.
One of the aims of plasmonics is to control the propagation of
surface plasmons by means of optical elements that could couple or
decouple light to surface
plasmons\cite{7,Weeber,MUGonzalez,Ilya1,Ilya2}, with the prospect of
developing a new technology consisting of
photonic nano-devices\cite{EkmelOzbay01132006,SPcircuitry,Zia}.
For this, the scattering properties of SPPs by typical configurations of scatterers
should be known. \\
In this article, we present a comparative systematic study of scattering of
surface plasmon polaritons by defects of different shapes. The
defects can be either indentations of the metal surface (grooves) or
protrusions on it (ridges). Some properties of ridges and grooves
have been investigated in several
works.\cite{Pincemin95,FJEbbessen,Sanchez-Gil(OSA),Sanchez-Gil(APL),FLT2005,Sondergaard}
Nonetheless, a clear general picture of what configurations are best
for different optical functionalities, has not yet emerged. This is
because plasmonic systems, such as ridges and
grooves, involve a large number of physical quantities at different
scales. As a consequence, dramatic changes may result from slight
variations of any of the many parameters involved: wavelength,
sizes, shapes, materials, the period in arrays of defects, types of
illumination and so forth.\\ We have previously presented a
comparative study of the scattering of \textit{shallow }ridges and
grooves\cite{Art1}, a case for which analytical expressions can be
obtained. In this paper we extend that study to consider the height
dependence of the scatterers, and focus on analyzing systematic
changes on the scattering properties, rather than on the
optimization of physical properties. We analyze the scattering of a
SPP by both individual defects and arrays of defects, in order to
elucidate how band-gaps effects affect the properties of individual
defects. In order to reveal more clearly the differences between
scattering properties of ridges and grooves, in this work we
consider the simplest case of bi-dimensional defects, which are
deemed infinite in one of the dimensions parallel to the interface.
 \section{\label{uno} Scattering Systems }
\begin{figure}
\includegraphics[scale=1,width=8cm]{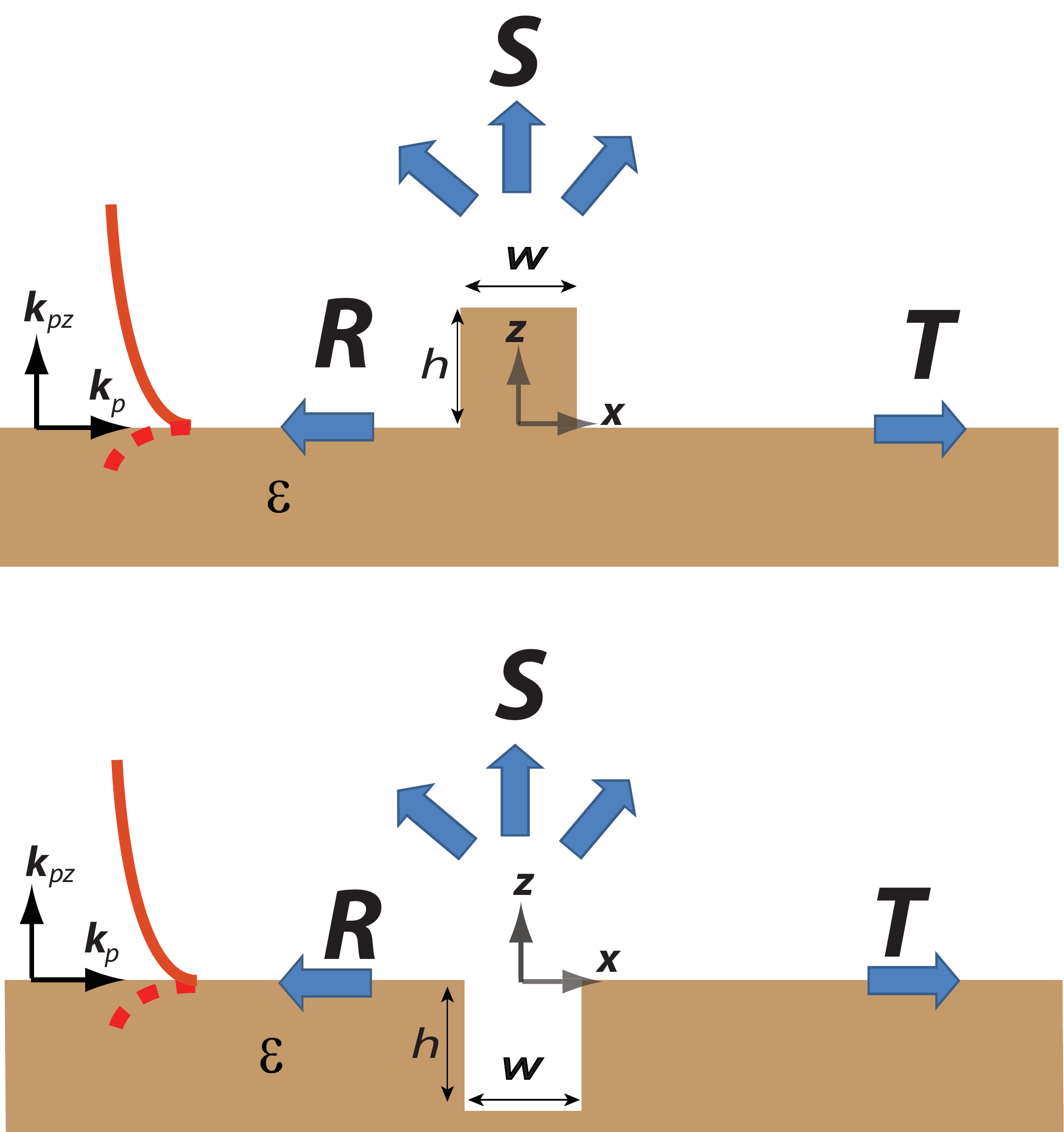}
\caption{Schematic representation of the two scattering systems: (a)
ridges (protrusions) and (b) grooves (indentations).
 A surface plasmon polariton impinges at normal incidence.
 Part of the incident energy flux of surface plasmon is transmitted,
 while another part is scattered in the reflection and out-of-plane radiation channels.
 In this paper the width of the defects is set to $w=100nm$. }\label{Scheme}
\end{figure}

Figure \ref{Scheme} represents a ridge and a groove and the
direction of the x and z axes.
   These systems are deemed
infinite in the direction perpendicular to the page.
The metal slab is considered to be optically thick, so the defect is effectively placed
on an air-metal interface.
\begin{figure}
\includegraphics[scale=1,width=8cm]{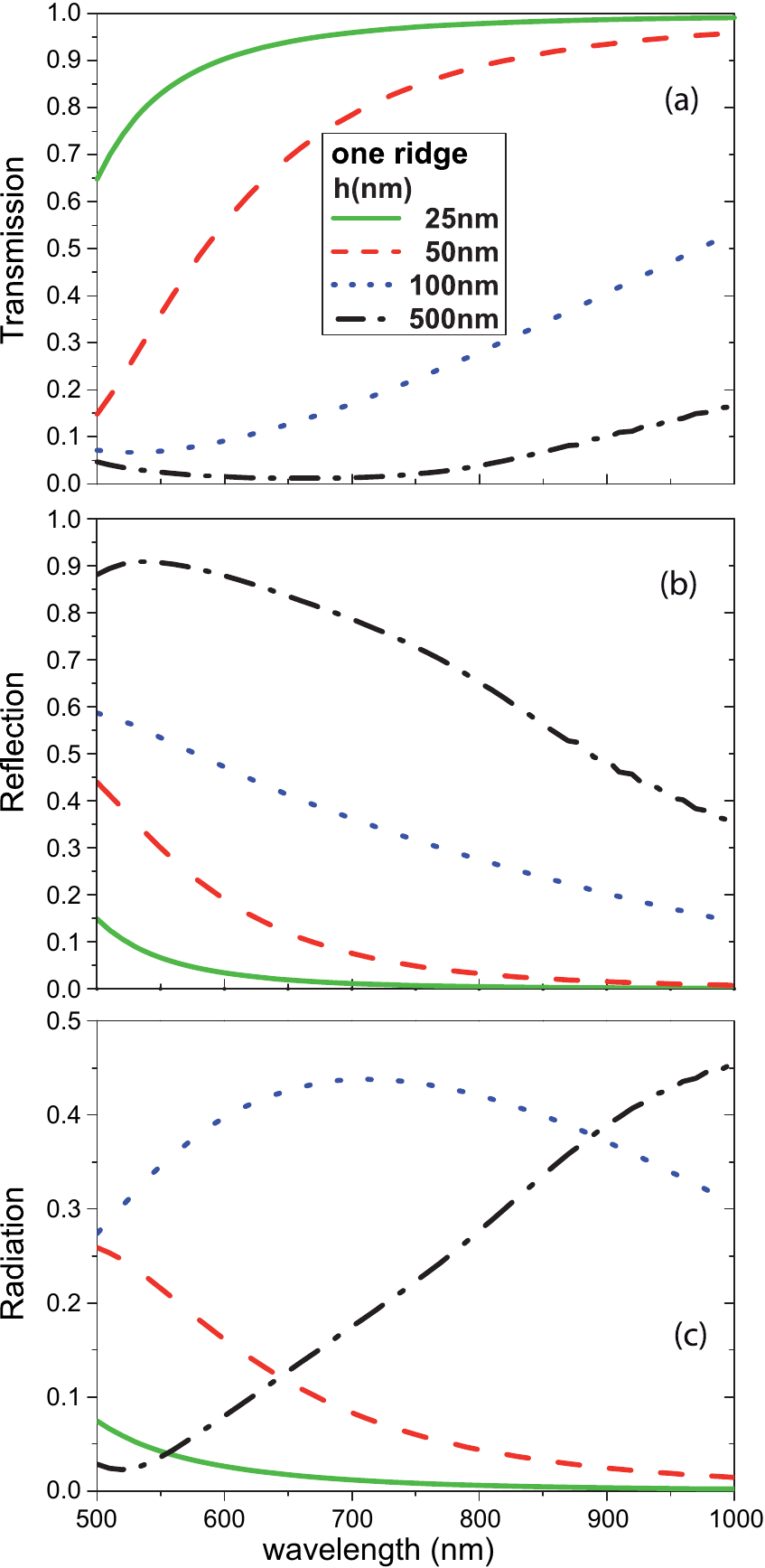}
\caption{ (Color online) The transmission (panel (a)), reflection (panel (b)) and
radiation (panel (c)) of a surface plasmon, as a function of
wavelength, produced by \textit{one ridge} at different heights. The
ridge width is fixed to $w=100nm$ while its height $h$ is varied from
$25nm$ to $500nm$. The material is silver.}\label{IndiRidge}
\end{figure}
 In all calculations throughout this paper, the width ($w$) of all defects is fixed to the
value of $w=100nm$ (which is experimentally viable). First, we shall
consider individual defects of several heights: from shallow defects
to deep defects. Secondly, we shall consider arrays of ridges and
grooves consisting of identical defects periodically distributed.
The considered material is silver and
its dielectric constant is taken
 from Ref.[\onlinecite{undici}]. Absorption is taken into account because the
size of the defects may be comparable with the SPP absorption
length. The solutions are attained through the Green tensor
approach, which is a standard theoretical method to solve
electromagnetic scattering problems
\cite{Hohmann,Protasio,Keller,Li,MGD,Felsen,NanoOptics,AlexImpedance}.\\
\indent Both ridges and grooves will be illuminated by a surface plasmon
at normal incidence, propagating in the x-direction with in-plane
wavevector\cite{Cottam,4} $k_p=(2\pi/\lambda)
\,\sqrt{\varepsilon\,(\varepsilon+1)^{-1}}$, where $\lambda$ is the
free-space wavelength.\\ The scattering problem is defined in terms
of the fraction of the SPP energy flux carried by
 Transmission (T), Reflection (R) and
out-of-plane Radiation (S). For economy of language, throughout the
paper we shall refer to the out-of plane radiation of the impinging
SPP energy flux simply as radiation.
  Unless otherwise
stated, transmission, radiation and reflection shall be represented
as a function of wavelength from a short-wavelength edge of visible
light at $\lambda_1=500nm$ to a long-wavelength edge of
near-infrared light $\lambda_2=1000nm$. The SPP has a penetration in
the air semi-space determined by $l_{air}=(\Im\{k_{pz}\})^{-1}$, where
$k_{pz}=2\pi/(\lambda\sqrt{\varepsilon+1})$ is the vertical
component of its wavevector.
 In the air-silver interface, the SPP penetration in air
 grows monotonically from visible to infrared wavelengths, from
the value of $l_{air}=\simeq 200nm$ at $\lambda_1$, to
the value of  $l_{air}\simeq 1080nm$ at $\lambda_2$.
Approximately at $700nm$, in the border between optical and infrared
wavelengths: $l_{air}\simeq 480nm$
\section{Individual Defects}
\subsection{Individual Ridges}
 Figure \ref{IndiRidge} renders the evolution of the SPP transmission
 spectrum along with the SPP reflection and radiation spectra,
  of an individual ridge as its height is varied from $25nm$ to $500nm$.\\
\begin{figure}
\includegraphics[scale=1,,width=8cm]{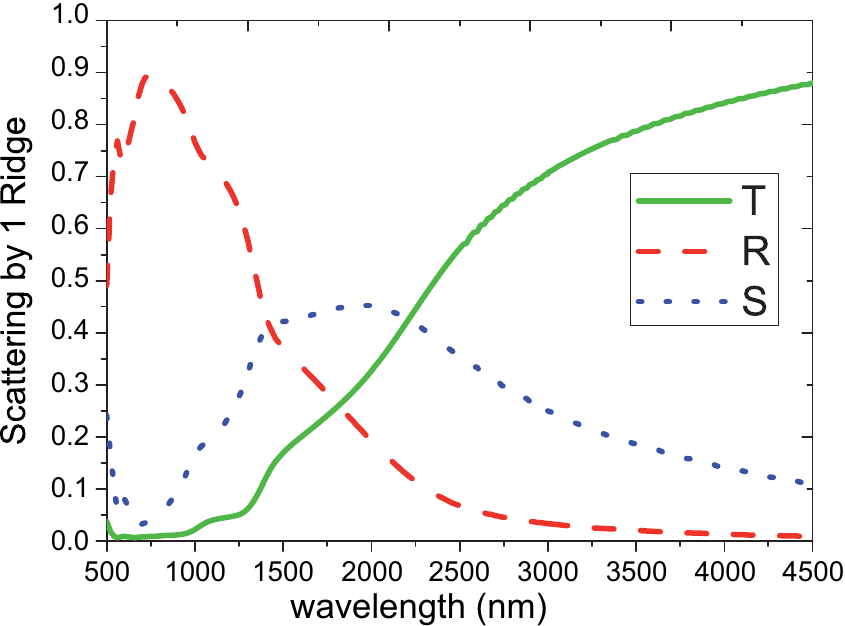}
\caption{(Color online) The transmission (T), reflection(R), and  radiation(S) of a
surface plasmon by one ridge as a function of wavelength. The ridge
width is $w=100nm$ while its height is $h=1500nm$. The material is
silver.}\label{FobaRidge}
\end{figure}
The scattering by shallow defects ($h<<\lambda$) has been previously
analyzed in Ref.[\onlinecite{Art1}]. In this case the size of the
defect is much smaller that the free-space wavelength of the
incident light and, therefore, the defect scattering can be
associated to the emission of a point-dipole. In 2D systems, such as shallow ridges
($h=25nm$), R and S exhibit a smooth Rayleigh-type decay with
wavelength, that scales as $\lambda^{-3}$.
Correspondingly, the transmittance for shallow defects increases monotonously
with wavelength. As shown in Fig. \ref{IndiRidge}(a) this monotonous decrease is
maintained in the case of a taller defect. The transmittance also decrease monotonously
with increasing height of the ridge, being very small ($T\sim 0.1$) when the defect height is
$h \sim l_{air}/2$.
As shown in Fig.\ref{IndiRidge}(b), the increase in
$h$ results in a monotone increase in reflection until the ridge is
almost a perfect reflector, reaching its maximum efficiency of
$90\%$ of the incident plasmon (including absorption) at optical wavelengths. \\
Figure \ref{IndiRidge}(c) shows the evolution of the radiation
spectrum for the same set of heights.  When the ridge is still
shallow ($h=\lambda_1/10$), the maximum radiation occurs at the
lowest wavelength considered. This maximum is red-shifted at every
height increment considered. Finally when the ridge height is
$h=500nm$ and $h>l_{air}$ in the optical range (up to $700nm$), the
radiation peak is red-shifted to near-infrared wavelengths
$\lambda>\lambda_2$. For all values of $h$ considered, the peaks in
the radiation spectrum
are smaller than the maxima in the reflection spectrum.\\
 Figure \ref{FobaRidge} represents an extremely tall ridge ($h=1.5\mu m$).
  This case is perhaps difficult to reach experimentally, but it is considered for academic reasons as
it contains information on the main scattering channels in different
regions of the spectrum. Notice that, this time we are representing
the spectra of T, R and S from a wavelength of $500nm$ to $4500nm$. In
this case, reflection is the the strongest scattering channel in the
optical region. Radiation, however, is the main scattering mechanism
in the interval [$1500-2000nm$] of the spectrum.  Even then,  and
unlike the situation in the reflection channel, radiation is not
systematically enhanced with the ridge height. In fact, its
scattering efficiency never overcomes $50\%$ of the incident energy
flux.
\begin{figure}
\includegraphics[scale=1,,width=8cm]{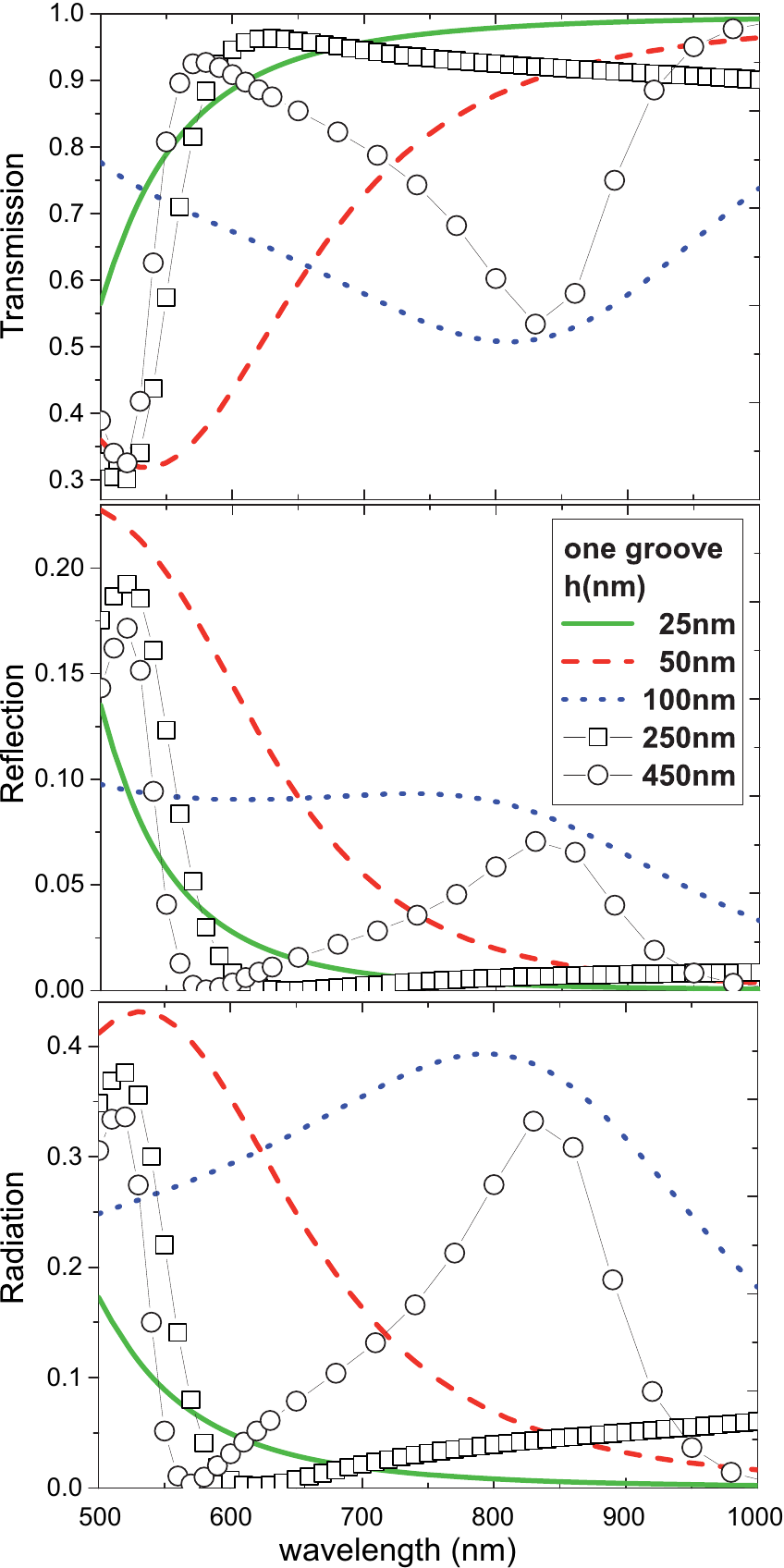}
\caption{(Color online) Scattering coefficients of a surface plasmon impinging onto a single groove:
transmission (panel (a)), reflection (panel (b)) and
radiation (panel (c)). The
groove width is fixed to $w=100nm$, but different groove depths have been considered, from $h=25nm$ to
$h=450nm$. The material is silver. }\label{IndiGroove}
\end{figure}
\subsection{Individual Grooves}
\begin{figure}
\includegraphics[scale=1,,width=8cm]{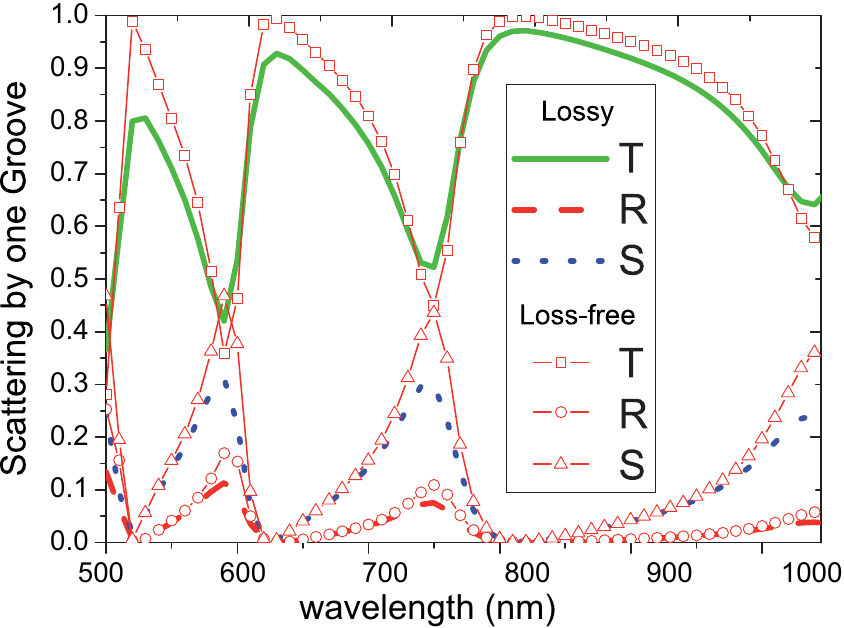}
\caption{(Color online)  Scattering coefficients of a surface plasmon impinging onto a single groove.
The groove width is $w=100nm$ while its height is $h=1000nm$. The material
is silver (in the loss-free case the imaginary part of the dielectric constant has been set
to zero).}\label{GrNoAbs}
\end{figure}

\begin{figure}
\includegraphics[scale=1,,width=8cm]{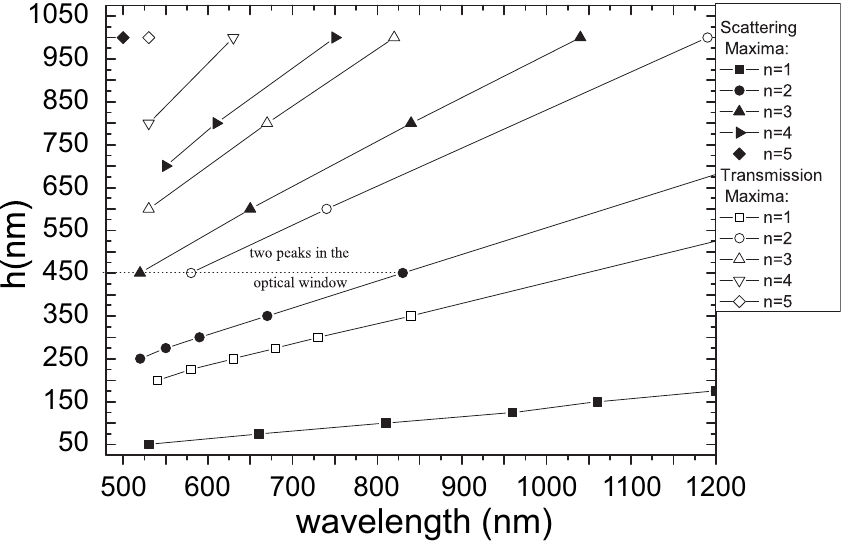}
\caption{The depths ($h$), of one groove of width $w=100nm$, against
wavelengths providing either maximum scattering or maximum
transmission of SPPs. The maxima, of both transmission and
scattering, that appear on the same straight line are labeled by an
integer index $n$. }\label{TMinima}
\end{figure}

 The surface plasmon scattering by a groove has been
 studied much more than the one by a ridge, see for example
Ref.[\onlinecite{FLT2005,M.Kuttge,Liu,Sanchez-Gil(OSA),Ioannis}].
  In a simple
model\cite{FLT2007,FJEbbessen} the field excited in the groove by
the incident SPP can be expressed as a superposition of the
waveguide modes, which propagate in the \textit{vertical
direction}. Therefore, resonances may arise due to reflection at the bottom
of the groove. Thus, the scattering cross section depends strongly with
$h$, presenting both maxima and minima (when the groove may be seen as a weak impedance defect\cite{AlexImpedance}, virtually invisible to the incident SPP). This is
illustrated in Fig.\ref{IndiGroove}. Let us first consider
the case of a shallow groove ($h=25nm$), which we have analyzed before\cite{Art1}.
In this case, the reflection and radiation coefficients present a Rayleigh-type
monotonous decay with wavelength. However, a resonant peak
appears close to $\lambda = \lambda_1= 500nm$ already for
$h=50nm$ (i.e. $h=\lambda_1/10$).
Let us label this resonance
$n=1$. Consider now $h=\lambda_1/5$. As a result of the depth
increase, the groove resonance $n=1$ is red-shifted and damped since
at at longer wavelengths transmission tends to
increase \cite{FLT2007}. \\
At a critical depth of a half-wavelength ($h=250nm$)
Fig.\ref{IndiGroove} exhibits  a new resonant wavelengths, let us
label it $n=2$.  When the groove depth is further increased
 the $n=2$ resonance is red-shifted (at $\simeq 850nm$) as seen for $h=450nm$.
 At the same time the optical path within the groove is large enough to meet
 a new resonant condition ($n=3$) in the neighborhood of
 $\lambda_1$. For greater depths this behavior culminates in
a multiple-resonance pattern, represented by a succession of damped
and broadened oscillation between points of maximum scattering and
maximum transmission as in Fig.\ref{GrNoAbs}.\\ In order to show the
effect of absorption on the system Fig.\ref{GrNoAbs} presents the
spectra for T, R and S, by a $1\mu$m deep groove with and without
absorption (in this case the calculation is done by setting the imaginary part of the
dielectric constant equal to zero). The figure shows that the spectral position of both the
resonant wavelengths and the transmission maxima are the same in the
loss-free and lossy case. Furthermore, in both cases, the net
scattering (reflection+radiation) is zero at wavelengths of
transmission maxima. The effect of absorption is to attenuate the
amplitude of the reflection and radiation peaks at resonant
wavelengths as well as those of the transmission maxima
(which in the loss-free case give T=1). \\
 As it turns out, the resonant
coupling of the groove and the plasmon results in both radiation and
reflection maxima, occurring at the same wavelengths. Yet, we find that
radiation is the most efficient scattering mechanism in a groove is
especially at long wavelengths, as seen for example
Fig.\ref{IndiGroove} and Fig.\ref{GrNoAbs}. \\
The position of the transmission and scattering maxima can be
calculated by making the assumption that the field propagates only
in the z-direction. This assumption implies that the relation
between wavelength and height at the critical points of maximum and
minimum scattering (or maximum transmission) is: $h=a \lambda +b $.
An estimate for $a$ and $b$ can be obtained by
neglecting the effect of the groove width on the resonant conditions
and considering that the groove is defined by walls made of perfect
electrical conductor (PEC).  This results in\cite{FLT2007,FJEbbessen}:
$a=a_{max}=(2n+1)/4$ and
$a=a_{min}={n}/{2}$ for for maximum and minimum scattering
respectively,  while $b=0$ in both cases.
\\ Our exact calculations are in accordance with the linear dependence
between $h$ and $\lambda$ at maximum and minimum scattering, as
shown in Fig.\ref{TMinima}. We have made linear fits on all
resonance and transmission curves in Fig.\ref{TMinima}, calculated
the relative error on them being straight lines (defined as the
ratio of the standard error on the slope and the fitted slope ). The
relative error is always smaller than $4\%$ in the optical range. We
have tabulated the slopes and intercepts of the linear fits of our
results, for the scattering maxima and minima in Table \ref{tab}.
\begin{table}
\caption{\label{tab} The table below presents the slopes and
intercepts of the linear fits of our results, for the scattering
maxima and minima.}
\begin{ruledtabular}
\begin{tabular}{ccccc}
 $n$ & $a_{min}$ & $b_{min}$ & $a_{max}$ & $b_{max}$\\
\hline 1 & 0.49 & -60.8  & 0.188  & -51.
\\
2 & 0.89 & -68.7& 0.63 & -74.\\
3 & 1.37 & -128 & 1.05&-91.4\\
4 & 2.& -260 & 1.48 & -113. \\
\end{tabular}
\end{ruledtabular}
\end{table}
The exact results, is consistent with the model of
Ref.[\onlinecite{FJEbbessen}] but shows deviations from it. This can
be associated to the penetration of the field in a real metal which
implies that the field inside a groove has a small wavevector
component parallel to the surface $k_x \ne 0$. Thus, while in a PEC
$k_z = 2\pi /\lambda$ in a real metal $k_z$ is larger.
Correspondingly, the condition of resonant depth occurs at smaller
values than those predicted by the PEC approximation. This is
reflected in the smaller values obtained for the coefficients
$a_{min}$ and $a_{max}$.
\begin{figure}
\includegraphics[scale=1,,width=8cm]{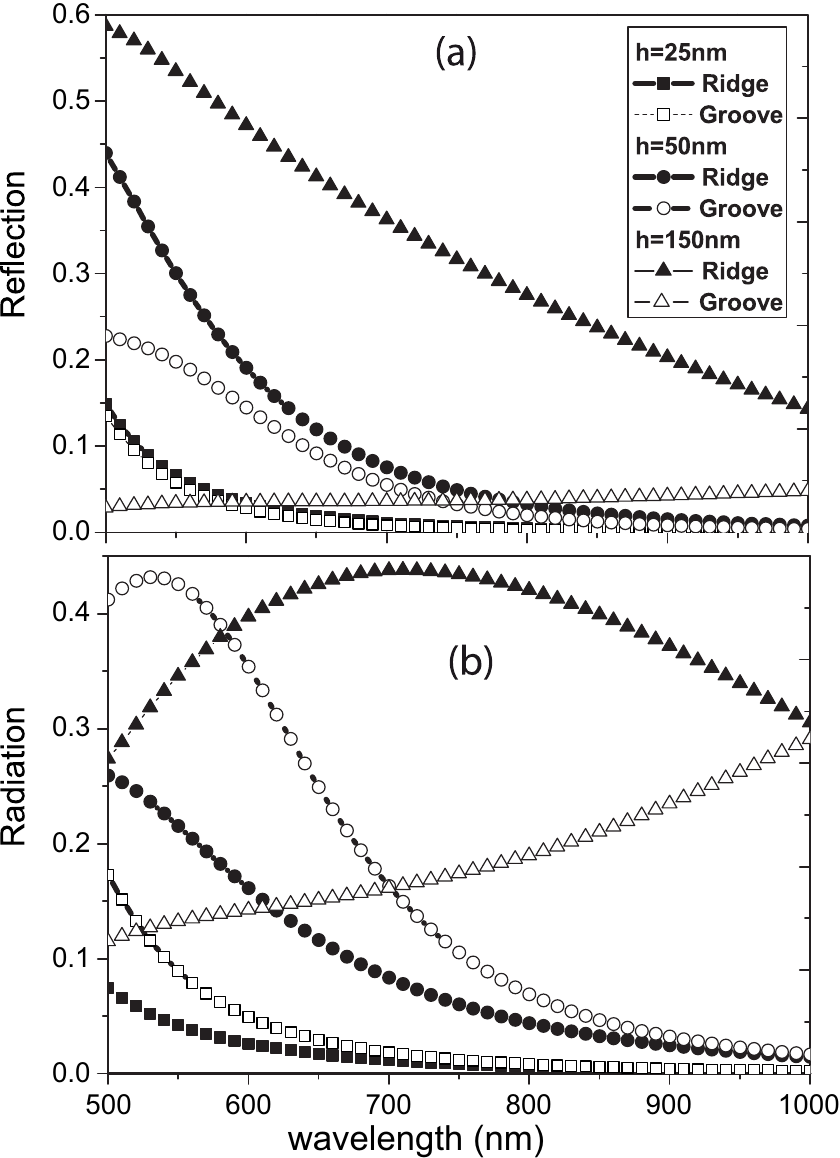}\\
\caption{ The reflection (panel(a)) and radiation (panel(b)) of a
surface plasmon polariton by both one ridge and one groove as a
function of wavelength, for different heights/depths ($h$). The
defects width is $w=100nm$ while their heights are$ h=25nm,50nm$ and
$150nm$. The material is silver. }\label{RvsG1}\end{figure}
\section{\label{Comparativa}Individual Ridges vs.
Individual Grooves: reflection and radiation}
\begin{figure}
\includegraphics[scale=1,,width=8cm]{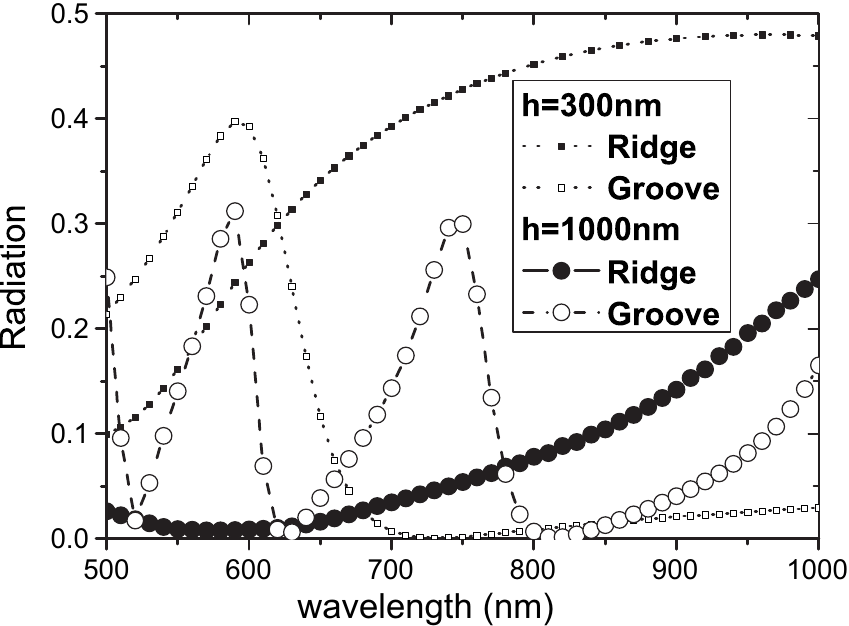}
\caption{The radiation of surface plasmon polaritons by both one
ridge and one groove as a function of wavelength, for different
heights/depths ($h$). The defects width is $w=100nm$ while their heights
are $h=300nm$ and $h=1000nm$. The material is
silver.}\label{ridvsGr3001000}
\end{figure}
So far we have seen that the ridge capability to reflect and radiate
SPPs increases systematically with height. In grooves, by contrast,
reflection and radiation of SPPs occurs mainly at resonant depths.\\
 In the Rayleigh limit the comparison between ridges and
groove, revealed two limiting cases\cite{Art1}: $i)$ When the
defects are very small and square, the reflection between ridges and
grooves of the same size is comparable, while the out of plane
radiation is always larger in a groove. $ii)$ When defects are
shallow but very long in the x-direction ridges and grooves have
similar scattering. Yet, this requires needle-type defects, with
$w>>h$ (at least $w>10h$). As the aspect ratio $(w/h)$ of the defect
is varied we pass gradually from the case $i)$ to the case $ii)$.\\
The scattering coefficients of the shallow defects considered
($w=4h$), can be qualitatively associated with case $i)$ as shown in
Fig.\ref{RvsG1}. The rest of the section is devoted to studying how
this scenario changes when the defects are not shallow.
\subsection{Reflection of surface plasmons}
Figure \ref{RvsG1}(a) compares the reflection spectra of ridges and
grooves of the same sizes, as $h$ is increased up to $h\sim
\lambda_1/3$. The figure illustrates the greater reflection
efficiency of ridges over grooves, for this range of values of $h$.
Ridges can efficiently reflect the incident SPP, even when their
height is only a few tens of nanometers. Figure \ref{RvsG1}(a) shows
that a $50nm$-tall ridge, at a wavelength $\lambda= 500nm= 10h$) is able to reflect
almost $45\%$ of the incident SPP. We find that, in agreement with
the results of Ref.[\onlinecite{M.Kuttge}],  a groove never
overcomes a reflection efficiency of $30\%$ of the impinging SPP
energy flux.
 For larger values of $h$ than those rendered in Fig.\ref{RvsG1}(a),
the ridge maximum reflection efficiency is much greater than that of
a groove with the same size and shape.
\subsection{Out-of-plane Radiation of surface plasmons}
The comparison between the radiation efficiency of SPPs by
grooves with the one by ridges of the same size, is more complex.\\
 Figure \ref{RvsG1}(b) shows that
 the groove is resonant at $\lambda=540nm$ for $h=50nm$.
 In a neighborhood of the resonant region, at optical wavelengths,
 the groove gives greater overall radiation
than a ridge with the same size.
 However when the groove is
non resonant ($h=150nm$), throughout the spectral region
$[\lambda_1,\lambda_2]$, its radiation is smaller than
the one by the ridge, in the entire spectral region.\\
A second resonance of the groove emerges at
$h=300nm$. As seen in Fig.\ref{ridvsGr3001000} such
groove resonance presents a localized peak over a relatively small
spectral region in the visible range. Instead, for a ridge of the
same height, the SPP radiation grows monotonously in the range
[$\lambda_1,\lambda_2$].
  Figure \ref{ridvsGr3001000} also shows what happens eventually
for very deep defects. The maximum of the ridge radiation spectrum
red-shifts and occurs at $\lambda>\lambda_1 $. The radiation
spectrum of a groove tends to become prominent in the optical
region, presenting a multi-resonant pattern of emission lines.

In conclusion, an individual ridge scatters the energy flux of an
incident SPP, more efficiently than a groove of the same height. The
related reflection and radiation spectra broaden as the ridge gets
taller (but still in the range $h \ll l_{air}/2$).
Radiation maxima occur at different wavelengths from
reflection maxima. However a single groove presents resonances that
depend on the groove depth. Resonant radiation and reflection peaks
occur at the same wavelength.
A resonant groove presents a larger radiation coefficient than a ridge of the
same size.

\begin{figure}
\includegraphics[scale=1,width=8cm]{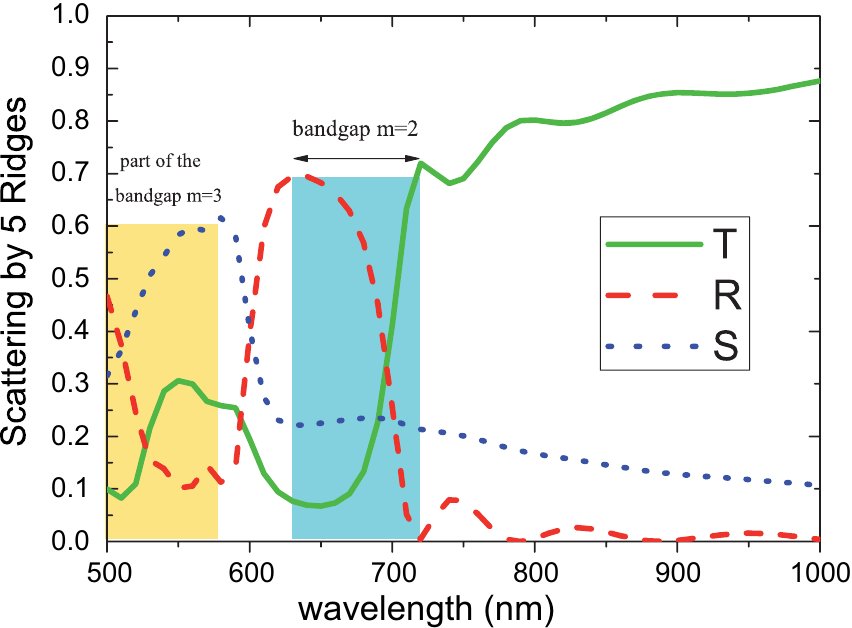}%
\caption{(Color online)  The transmission (T), reflection(R), and  radiation(S) of a
surface plasmon by an array of five ridges as a
function of wavelength. The period is $600nm$, and each ridge has width  $w=100nm$ and height
$h=50nm$. The material is silver. The spectral positions of the
band-gaps are indicated.}\label{Ridges50}
\end{figure}

\section{Arrays of surface defects}
We shall now highlight the effects exhibited collectively by
arrays of either ridges
or grooves, which do not appear in individual defects, and then
comment briefly on the cases in which scattering is explainable in
terms of the individual behavior of defects.\\
 As known, a
set of scatterers periodically distributed, exhibits photonic
band-gaps\cite{BG,BK}. We have considered arrays of five
\textit{identical defects} of width $w=100nm$, with a periodicity of
$600nm$, \textit{and with height $h$, which is varied}. As it turns
out, this is a sufficient number of defects to observe band-gaps
effects. The periodicity is chosen
to produce band-gap effects within the optical range.\\
 Most of the phenomenological analysis, for our scattering system, is based on the
concept of band-gap structure.
  The position of the band-gap short-wavelength edge
$\lambda^{(m)}_{-}$ is determined by the Bragg
law\cite{BK,FLT2005,MUGonzalez} $k_p d=m\pi$.
 The width of the band-gap
spans from $\lambda^{(m)}_{-}$ to $\lambda^{(m)}_{+}$. Former
results\cite{FLT2005,Sondergaard} have shown that reflection has a
peak at $\lambda^{(m)}_{-}$ while radiation has a peak at
$\lambda^{(m)}_{+}$.\\
 The short-wavelength edge of the band-gap $m=2$ is at about
$\lambda^{(2)}_{-}\approx 630nm$. This is the main band-gap
observable in our system within the spectral range
$[\lambda_1,\lambda_2]$.
  Yet the $m=3$ bands-gap
will be part of the discussion.
\begin{figure}
\includegraphics[scale=1,width=8cm]{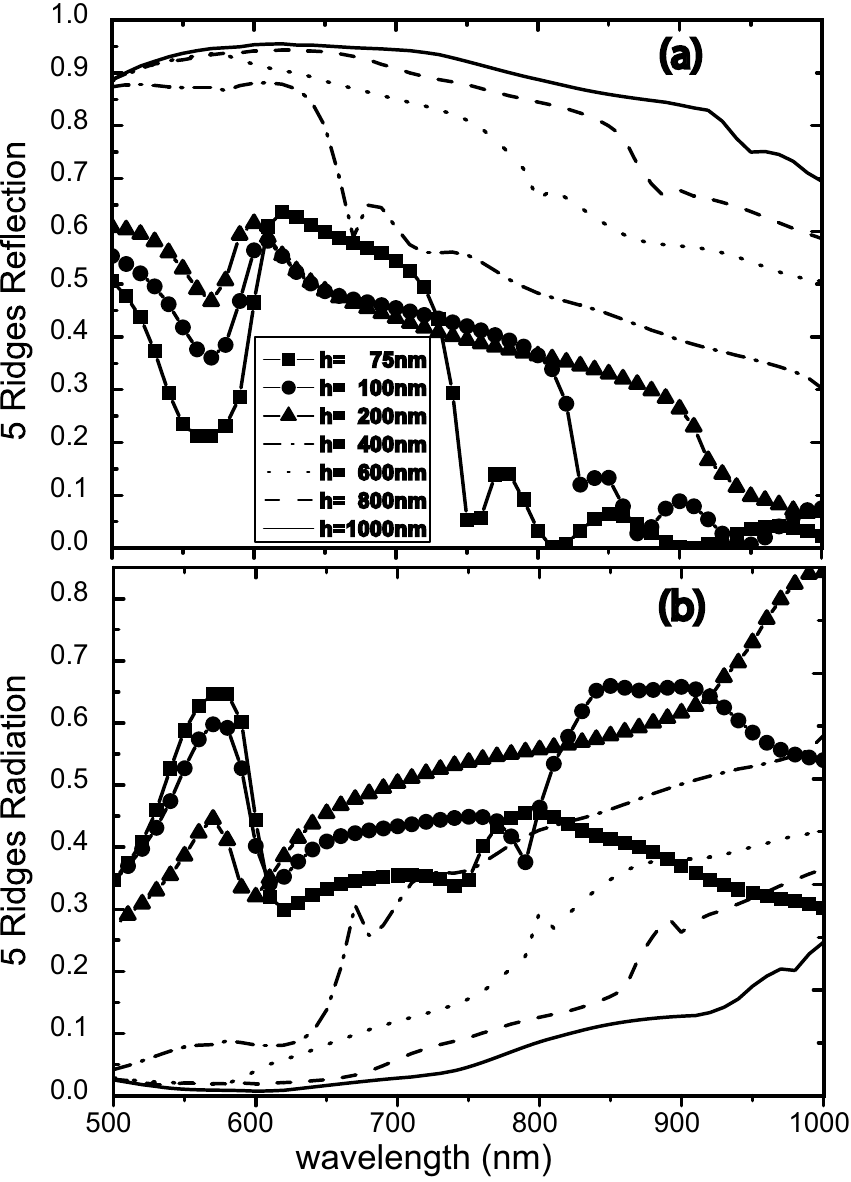}%
\caption{(Color online) The reflection (Panel(a)) and radiation (Panel(b)) of a
surface plasmon polariton by an array of five ridges (period $600nm$)
as a function of wavelength, for different heights ($h$). Each ridge
has width  $w=100nm$ while its height is varied from $h=75nm$ to
$h=1000nm$. The material is silver. }\label{RidgesR251000}
\end{figure}
\subsection{Ridges arrays: Collective effects }
For the chosen period, ridge arrays produce band-gaps that span a
region within $[\lambda_1,\lambda_2]$. In Fig.\ref{Ridges50} we can
notice the whole $m=2$ band-gap
$[\lambda^{(2)}_{-},\lambda^{(2)}_{+}]$ as well as  the
long-wavelength edge $\lambda^{(3)}_{+}$ of the $m=3$ band-gap at
$550nm$. At wavelengths in the neighborhood of $ \lambda^{(3)}_{+}$
the impinging SPP energy is scattered mainly in the radiation
channel, while in a neighborhood of $ \lambda^{(2)}_{-}$, it goes
mainly into the reflection channel. These two band-gaps are very
close together. When the incident SPP free-space wavelength is just
outside the $m=3$ band-gap $ \lambda> \lambda^{(3)}_{+}$, the main
scattering mechanism becomes the build-up to the reflection maximum,
up to the wavelength $ \lambda^{(2)}_{-}$.  Notice that the $m=3$
band-gap is opened by fulfilment of Bragg's condition at about
$430nm$. Yet we have not considered this region because the SPP
propagation in silver, at these wavelengths, is curtailed by large
absorption effects.\\
An increase in the height ($h$) of the ridges in the array results in
band-gap broadening (see Fig.\ref{Ridges50}). As the band-gap $m=2$ gets larger,
the reflection peak at $ \lambda^{(2)}_{-}$ becomes more asymmetric. In fact, for a
fixed $h$, the asymmetry in the reflection peak $ \lambda^{(2)}_{-}$
is caused by the progressive growth of the radiation spectrum
throughout the gap. The band-gap, and hence the growth of the
radiation spectrum, spans from $\lambda^{(2)}_{-}$ to
$\lambda^{(2)}_{+}$. If, in a different configuration,
$\lambda^{(2)}_{+}$ is shifted to a longer wavelength
${\lambda'}^{(2)}_{+}$, the related radiation growth is also
extended to ${\lambda'}^{(2)}_{+}>\lambda^{(2)}_{+}$, resulting in a
more asymmetric reflection peak at $\lambda^{(2)}_{-}$.
\textit{Therefore the asymmetry of the reflection peak is also
associated to the spectral width of the gap.} Figure
\ref{RidgesR251000}(a) shows that if $h$ is increased from $h=75nm$ to
$h=200nm$, the resulting band-gap expands its width to include a
broader spectral region.
\\Notice
that the reflection spectra in Fig.\ref{RidgesR251000}(a), exhibits
a set of small multiple-resonance peaks at  $\lambda>750$nm, for
$h=75nm$ and $h=100nm$ . Correspondingly, radiation also exhibits such
small peaks, in the near-infrared region of the spectra rendered in
Fig.\ref{RidgesR251000}(b).  The small peaks are array finite-size
effects formerly discussed in Ref.[\onlinecite{Pincemin95}]. Notice
that, as explained in Ref.[\onlinecite{Pincemin95}], finite-size
effects occur outside the band-gap edges.
\begin{figure}
\includegraphics[scale=1,,width=8cm]{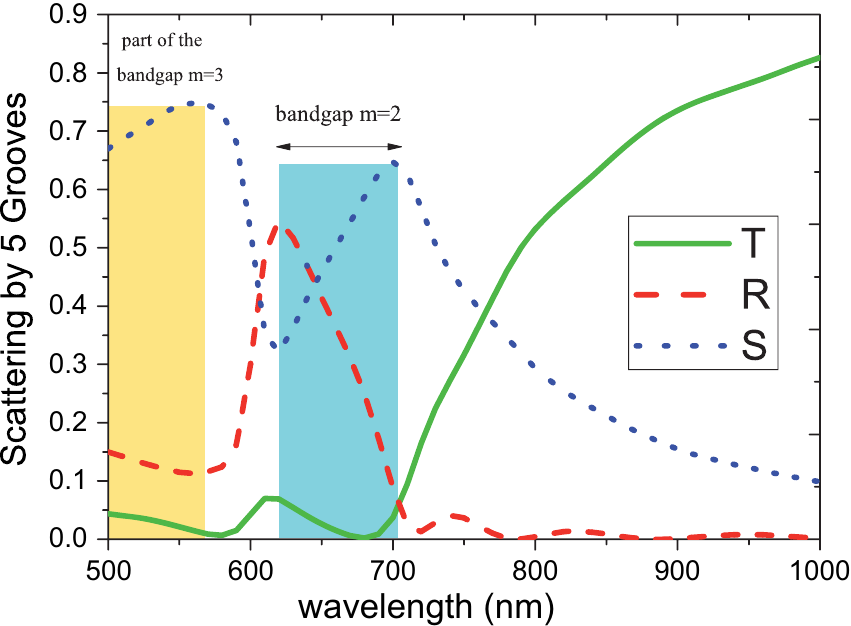}
\caption{(Color online) The transmission (T), reflection(R), and  radiation(S) of a
surface plasmon by an array of five grooves, and period $600nm$ as a
function of wavelength. Each groove has a width  $w=100nm$ and depth
$h=50nm$. The material is silver. The spectral position of the
band-gaps is indicated.}\label{Grooves50}
\end{figure}
 For $h=400nm$ the
height of the ridges is such that $h \sim l_{air}$, in the
range $\lambda\in[500-600]nm$. At these wavelengths such arrays
consist of ridges so tall that the first one or two block most of
the impinging plasmon. As seen in
Fig.\ref{RidgesR251000} when $h$ is increased from 400 to
$1000nm$, reflection rapidly becomes the only scattering channel in the range
 $[\lambda_1,\lambda_2]$ while radiation vanishes. Therefore, as $h$ approaches $1000nm$,
band-gap effects tend to disappear in the whole optical spectrum and
the scattering of the incident SPP by the ridge array, can be
interpreted, mainly, as the individual scattering of the first one
or two ridges.
 \subsection{Grooves arrays: Collective effects}
\begin{figure}
\includegraphics[scale=1,,width=8cm]{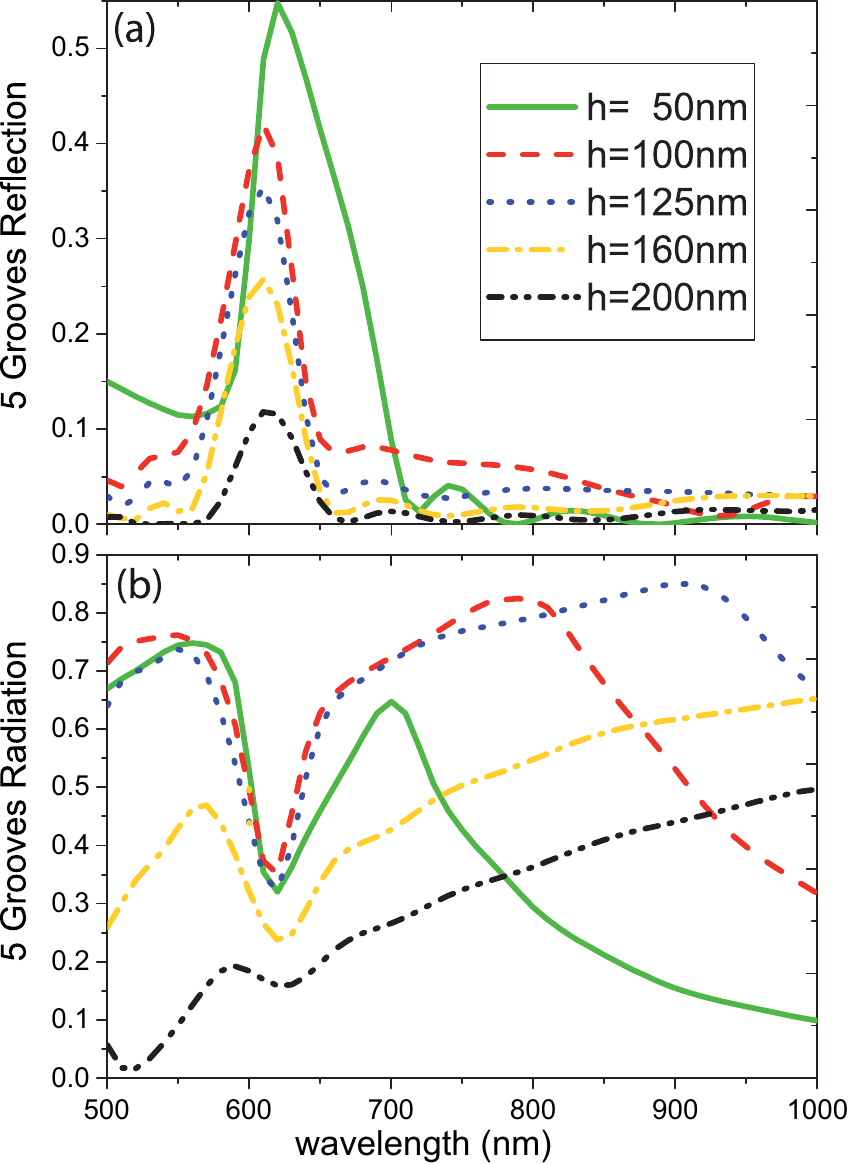}%
\caption{(Color online) The reflection (Panel(a)) and radiation (Panel(b)) of a
surface plasmon polariton by an array of five grooves, and period
$600nm$, as a function of wavelength, for different depths ($h$). Each
groove has width $w=100nm$ while its depth is varied from $h=50nm$ to
$h=1000nm$. The material is silver.}\label{GroovesR50200}
\end{figure}
\begin{figure}
\includegraphics[scale=1,,width=8cm]{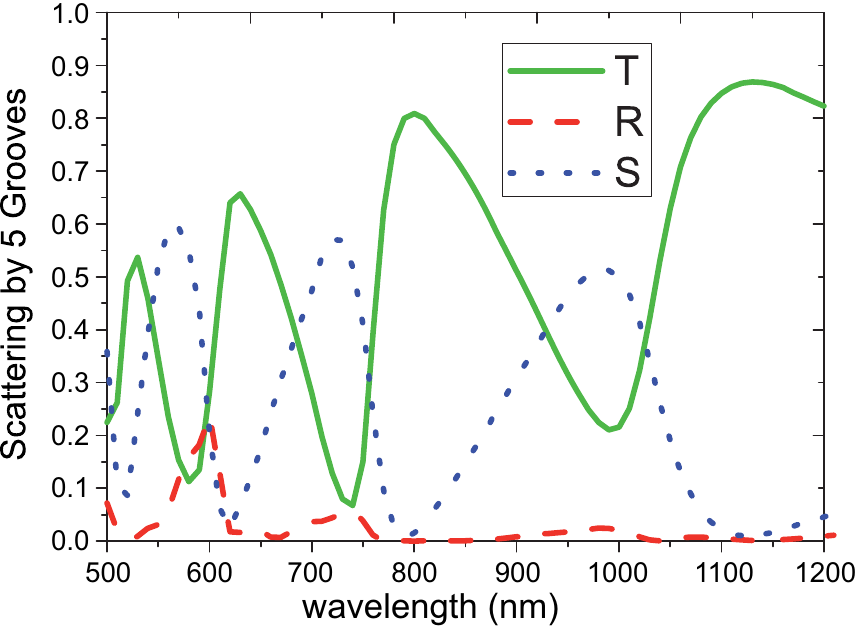}
\caption{(Color online) The transmission (T), reflection(R), and  radiation(S) of a
surface plasmon by an array of five grooves, and period $600nm$, as a
function of wavelength. Each groove has a width  $w=100nm$ and depth
$h=1000nm$. The material is silver.}\label{Grooves1000nm}
\end{figure}

As found for individual grooves, the scattering coefficients of
grooves arrays have an oscillatory behavior that depends on the
depth of the grooves and the free-space wavelength of the incident
SPP. Several works \cite{Sanchez-Gil98,Sanchez-Gil(APL),FLT2005,FLT2007,FJEbbessen,PhysRevLett.100.123901} have studied
SPP scattering by groove arrays.  In
order to compare with the ridge array analyzed before, in the following, we shall consider the SPP scattering by 5 grooves with
period $600nm$ and width $100nm$, and different depths. \\
First of all in Fig.\ref{Grooves50} we present the exact result for the array of grooves
considered in Ref.[\onlinecite{FLT2005}] with an approximate method.
 The two results are extremely
similar, except that the radiation growth within the band-gap is
much more evident in the exact result than in the approximate one.
Noticeably Fig.\ref{Grooves50} features the same radiation peak at
$\lambda_+(3)$ as Fig.\ref{Ridges50}.
\\ As opposed to the individual behavior of a resonant groove,
 in groove arrays the reflection and radiation maxima are effectively
 decoupled,
  when the grooves are interacting collectively.
 In fact, in this case the reflection peak
is imposed at $\lambda_-^{(m)}$ by Bragg's interference, based on
the period of the array. However, since the radiation peak is caused
by the decrease of the reflection (and transmission) within the gap
the two peaks appear at different wavelengths, unless the spectral
width of the band-gap is zero.
\\   When, in the same
configuration, the depth of the grooves is increased up to $h=200nm$,
the interaction between the incident SPP and the groove weakens. As a result
 the related reflection maximum is systematically reduced, as seen in
Fig.\ref{GroovesR50200}(a). Figure \ref{GroovesR50200}(b) shows that
the radiation maximum at $\lambda_+^{(3)}$ disappears.  Similarly,
as $h$ increases, the radiation also vanishes throughout the spectrum
$[\lambda_1,\lambda_2]$. The radiation peak at $\lambda_+^{(2)}$ is
only observable for $h=50nm$ and $h=100nm$, while for deeper grooves it
is red-shifted beyond
$\lambda> \lambda_2$.\\ 
Figure \ref{Grooves1000nm} represents the result for very deep
grooves. Note the small reflection peak at $\lambda_-^{(2)}$
is the only one decoupled from radiative emission lines.
 Comparison with Fig.\ref{GrNoAbs} suggests that the
interaction between the 5 grooves in Fig.\ref{Grooves1000nm} is weak
and that the scattering produced by the array is rather a
superposition of the individual behavior of each groove. The
comparison between the two figures also shows that the values of
radiation peaks are about 60\% in Fig.\ref{Grooves1000nm} and about
30\% in Fig.\ref{GrNoAbs}. That is, at resonance, the fraction of
impinging SPP flux radiated by five grooves  doubles the one
radiated by a single groove, of the same size. Besides radiating
more efficiently, five grooves produce more dissipation of the
impinging SPP energy flux than one does, and this causes less overall
transmission throughout the spectrum in Fig.\ref{Grooves1000nm}.\\
Finally notice that, as found for individual grooves, the
out-of-plane radiative loss is the dominant scattering mechanism for
groove arrays. However arrays of grooves can produce considerable
reflection of surface plasmons.
\section{Arrays of Ridges vs Arrays of Grooves}
As seen, ridges produce larger reflection efficiencies than grooves,
 even when the latter are resonant.\\
Comparing the increase of the radiation peak at $\lambda^{(2)}_{+}$
in Fig.\ref{Grooves50} with that in Fig.\ref{Ridges50} suggests that
grooves are better radiative emitters when the defects are shallow.
Accordingly, in groove arrays the main scattering mechanism within
the band-gap is radiation. In ridge arrays the main scattering
mechanism  within the band-gap is reflection. The spectral width of
the band-gap produced by either ridges or grooves of $h=50nm$ is
practically equal. Note that grooves are not resonant within the band-gap
$[\lambda^{(2)}_{-},\lambda^{(2)}_{+}]$ for $h=50nm$ (see
Fig.\ref{TMinima}). However, for $h=100nm$ grooves are resonant
within the spectral range $[\lambda^{(2)}_{-},\lambda^{(2)}_{+}]$.
In this case the band-gap width produced by the groove array is
larger than the one produced by the ridge array. In fact, in grooves
$\lambda^{(2)}_{+} \approx 900nm$ (see Fig.\ref{GroovesR50200}(b)), while in ridges
$\lambda^{(2)}_{+}\approx 800nm$ (see Fig.\ref{RidgesR251000}(a)).\\
Deep groove arrays undergo larger dissipative loss than tall ridges, and their maximum
out-of-plane scattering efficiency is about 60\%. Tall ridges
dissipate less of the SPP energy, achieving a maximum reflection
efficiency over 90\%.
\section{Conclusions}
We have studied the individual and collective scattering of ridges
and grooves in the optical range. The width of the defects was
always fixed to a typical value of $100nm$ and the period in arrays
was fixed to $600nm$, while the height and depths were varied from
$25nm$ to about one micron. To the best of our knowledge this is the
first comparative treatment between ridges and grooves where their
depth is systematically increased.  We found ridges are very good
reflectors, featuring (90\%) reflection efficiency, including
absorption. The related reflection becomes the main scattering
channel in the optical range as the ridge height is increased, while
radiation is red-shifted to infrared wavelengths. The reflection and
out-of-plane radiation maxima are found at different wavelengths.
Ridges produce more scattering than groves in general, but the
latter are more versatile. In fact, by adjusting their depth to the
free space wavelength, we can produce a tunable resonance or virtual
invisibility. At resonance grooves exhibit radiation and reflection
at the same wavelength, but these can be decoupled through band-gap
effects. Reflection is the least efficient mechanism for both
individual and collections of grooves. In both cases the reflection
peaks tend to disappear in the long-wavelength limit.\\ In shallow
arrays ($h<50nm$) ridges and grooves have similar band-gap structures.
However, within the gap, ridges can reflect incident SPPs more
efficiently than grooves. In turn, grooves can radiate the incident
SPP more efficiently than ridges. An array consisting of grooves
whose depth is resonant at wavelengths within the band-gap, produces
a larger band-gap spectral width than an array of ridges with the
same size and period.\\ A SPP suffers large dissipative losses when
scattered by a deep groove array, and the maximum fraction of the
impinging SPP energy flux scattered by a groove array is 60\%. A
tall ridge array can reflect more than 90\% of the impinging SPP
energy flux.
  \section{Acknowledgments}
The authors acknowledge financial support from the Spanish Ministry
of Science and Innovation under grants MAT2008-06609-C02,
CSD2007-046-Nanolight.es and NO.AP2005-5185.

\end{document}